\documentclass[preprint, 12pt]{elsarticle}
\usepackage{epsfig}
\usepackage{amssymb}
\usepackage{rotating}

\begin{document}

%
\begin{frontmatter}

\title{Constraints on Inelastic Dark Matter Signal using ZEPLIN-II Results}

\author{D.B. Cline}
\author{W. Ooi\corref{cor1}}
\cortext[cor1]{Corresponding author; address: Department of Physics \& Astronomy, University of California, Los Angeles, USA}
\ead{ooi@ucla.edu}
\author{H. Wang}
\address{Department of Physics \& Astronomy, University of California, Los Angeles, USA}

\begin{abstract}
There has been an increasing interest on the concept of Inelastic Dark Matter (iDM) - motivated in part by some recent data. We describe the constraints on iDM from the results of the two phase dark matter detector ZEPLIN-II, which has  demonstrated strong background discrimination capabilities ($>$98\%). Using previously published estimates of the ZEPLIN-II residual background, the iDM limits presented here exclude a significantly larger iDM parameter space than the limits derived without background subtraction, suggesting the exclusion of iDM signal claims at $>$99\% C.L., for Weakly Interacting Massive Particles (WIMPs) masses $>$100 GeV.  
\end{abstract}

\begin{keyword}
ZEPLIN-II \sep dark matter  \sep WIMPs \sep liquid xenon \sep radiation detectors 
\PACS 95.35.+d \sep 14.80.Ly \sep  29.40.Mc \sep  29.40.Gx
\end{keyword}

\end{frontmatter}

%
%
A liquid xenon (Xe) detector with two phase readout structure is well suited to the search for dark matter~\cite{cline2000}~\cite{hanguo}. This is due in part to the high atomic  mass of Xe and in part to the powerful background rejection ($>$98\%) of the two phase detector, which was first applied to the dark matter search by the UCLA Torino group in 1996. We show here the key steps on how the background was measured for the ZEPLIN-II detector, as were first reported in Ref.~\cite{zepsi},  and applied the recent results to place constraints on Inelastic Dark Matter (iDM) model.
\section{ZEPLIN-II Operation}
ZEPLIN-II is a dual phase (gas and liquid) Xe detector designed specifically to identify nuclear recoils induced on Xe atoms by the Weakly Interacting Massive Particles (WIMPs); one of the popular candidates for dark matter in the universe~\cite{turner},~\cite{jungman}. The detector is located in the UK Boubly underground laboratory which provides about 2800 m water-equalvent shielding for a 10$^{6}$ reduction in cosmic muon flux~\cite{muon}. The central detector was constructed out of low-radioactive materials at UCLA, with underground ambient radioactivity carefully studied and modeled at Boulby~\cite{bungau}. Additional background rejection is afforded by the dual phase design which measures both scintillation and ionization components for each event in the liquid xenon. Since background electron recoils differ from the signal nuclear recoils in the ratio of these components by about a factor of 3, background rejection can be achieved at $>$98\% efficiency~\cite{zepsi}. ZEPLIN-II detector took data for 31.2 days with 7.2 kg fiducial mass and obtained the final data set reported in Ref.~\cite{zepsi}. The WIMP signal acceptance box is defined between 5 to 20 keV electron-equivalent (keV$_{ee}$), encompassing 50\% of the total signal region determined by prior calibration runs. 29 events were observed in the acceptance box at the end of first dark matter run. However, since the acceptance box is flanked on both sides by background events, it is to be expected that some of the events in the box come from the tail distributions of the background events. A careful analysis of the background distribution was performed, and it was learned that 28.6 $\pm$ 4.3 background events were expected in the acceptance box. The detailed analysis and results are documented in Ref.~\cite{zepsi}, and are summarized here.

%
\section{ZEPLIN-II Background Measurement}
The background events in the S2/S1$\sim$300 region above the acceptance box are $\gamma$ events, as learned from calibration runs with $\gamma$ source $^{60}$Co and americium/beryllium (AmBe) neutron source. To quantify the event rate distribution, events with energy between 5 to 20 keV$_{ee}$ were selected and binned according to its Log(S2/S1)-k(E) values, where k(E) is the 50\% nuclear recoil acceptance value of log(S2/S1) for the energy of each event. In another words, events with Log(S2/S1)-k(E)$<$0 reside in the acceptance box. The $\gamma$ event rate distribution is well characterized by a Gaussian with an offset, where the offset accounted for the coincidental events arising from the high trigger rate ($\sim$70 Hz) encountered during calibrations. These coincidental events were not expected to contribute to the background level in the acceptance box, since they were not observed in the science data, which occurred at a much lower trigger rate ($\sim$2 Hz). As such, only the Gaussian component of the fit is used in background estimation, where the Gaussian is scaled to the overall event count in the science data and integrated up to the 50\% nuclear recoil boundary. In total, 16.1 $\pm$ 2.9 $\gamma$ events were estimated to be in the acceptance box between 5 to 20 keV$_{ee}$.\\
The second background events found below the acceptance box were nuclear recoil events with diminished S2 originated from radon contamination in SAES getter, as confirmed by an dedicated Rn measurement. These recoils occurred near the charged PTFE walls which stripped some of the charges from the electron cloud that constituted S2. The diminished S2 subsequently lowered the position reconstruction accuracy, which resulted in a fraction of these events being wrongly placed within the fiducial volume. To quantify the number of misplaced events, a distribution of reconstructed radii of all events within the acceptance regions in (S2/S1)-energy parameter space was studied. This distribution is well fitted with a Gaussian and is extrapolated into the fiducial volume. Finally the extrapolated distribution is  integrated to give the expected number of radon-related events, which is 12.5 $\pm $ 2.3. Refer to Table~\ref{expectation} for a summary of background estimation.
\begin{table}[h!]
\centering

\vspace{5mm}
\begin{tabular}{c|c|c|c|c}
Energy Range & Observed  & $\gamma$-ray ($^{60}$Co)  & Rn-initiated & Total \\
\hline
5-10~keV$_{ee}$ & 14  & 4.2 $\pm$ 2.4 & 10.2 $\pm$ 2.2 & 14.4 $\pm$ 3.3\\
10-20~keV$_{ee}$ & 15 & 11.9 $\pm$ 2.7 & 2.3 $\pm$ 0.5 & 14.2 $\pm$ 2.7\\
\end{tabular}
\caption{Overall expectation values in the nuclear recoil acceptance window compared to observed counts. The errors are derived directly from fit uncertainties.\label{expectation}}
\end{table} 

%
\section{Inelastic Dark Matter Model}
\begin{figure}[h!]
\begin{center}
$\begin{array}{c@{\hspace{.01in}}c@{\hspace{.01in}}c}
\includegraphics[width=3.1in]{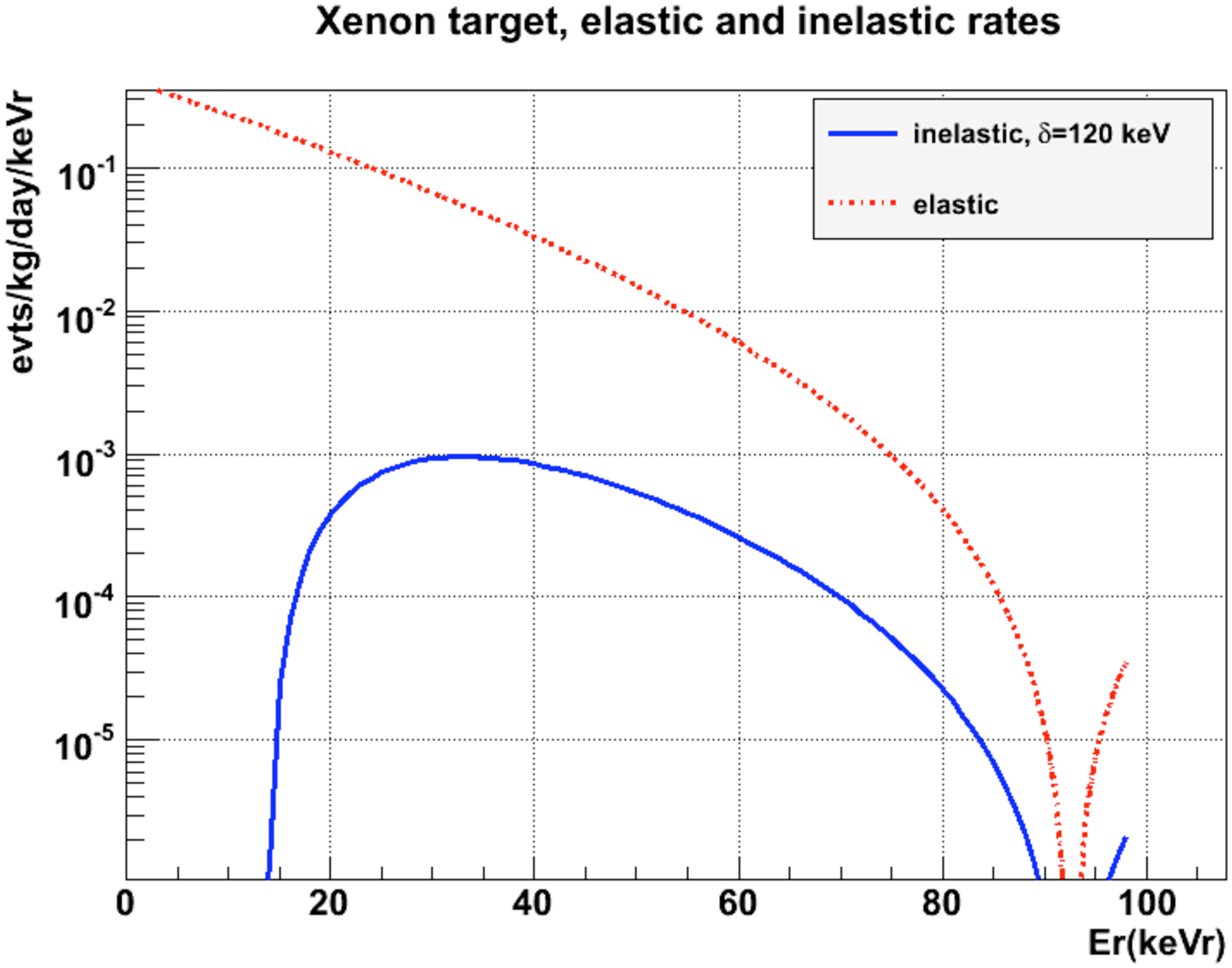} \\  
\includegraphics[width=3.1in]{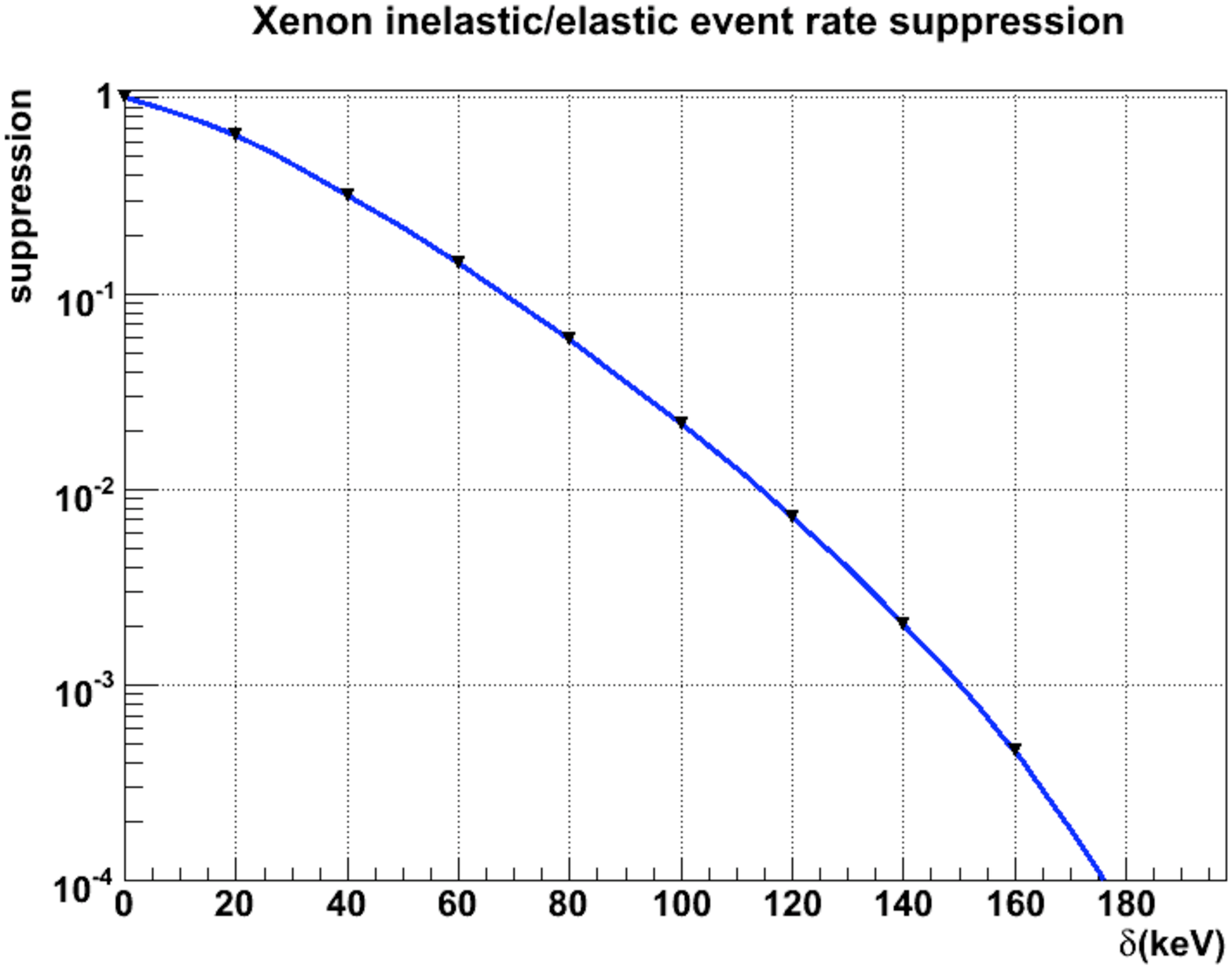} \\
\includegraphics[width=3.1in]{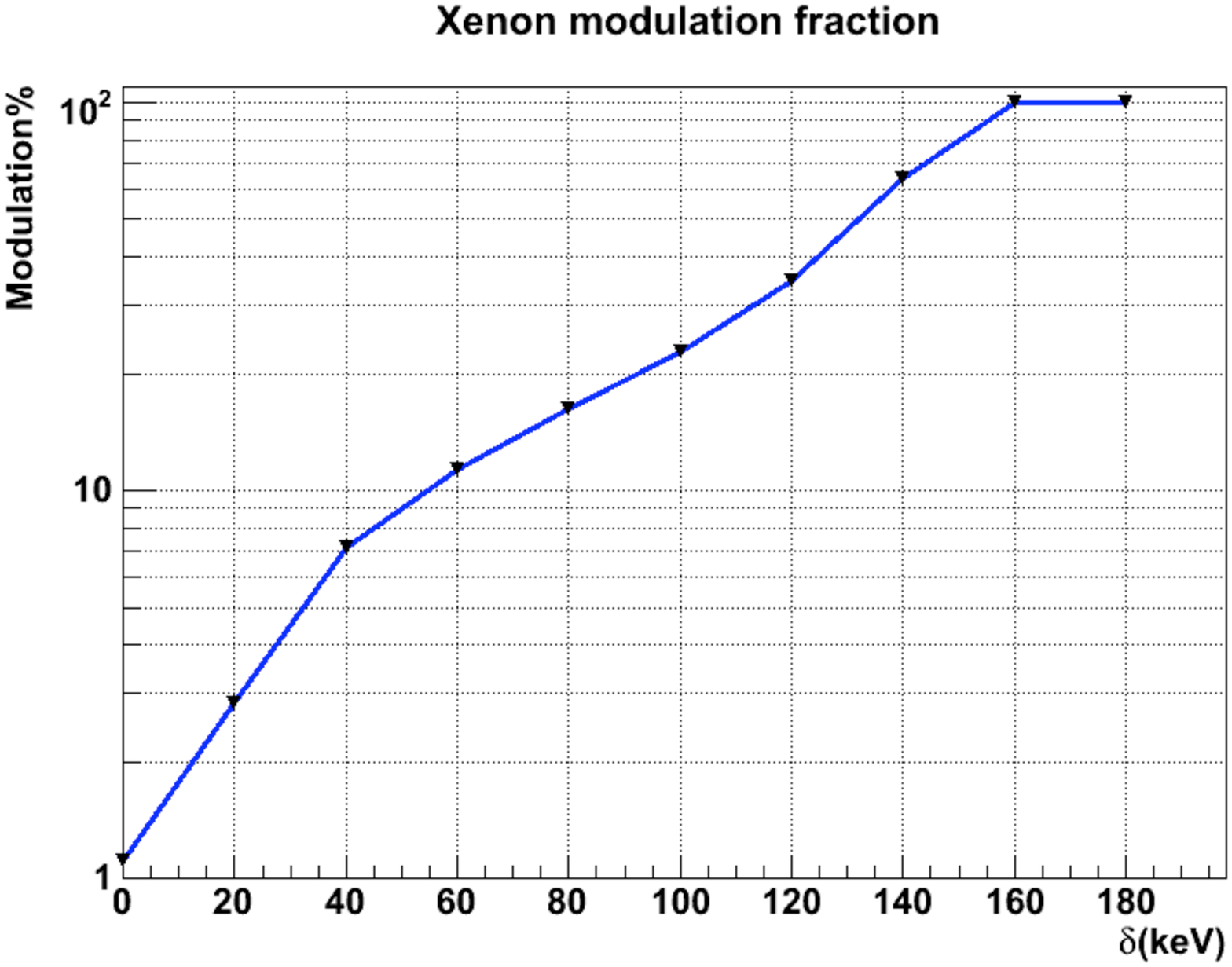} 
\end{array}$
\caption{\footnotesize{(top) Energy spectrum of WIMP events on Xe target for both elastic scattering
(red, dotted) and inelastic scattering (blue, line) models. (middle) Inelastic event rate suppression, compared to the elastic case, as a function of the splitting $\delta$. (bottom) Xe annual modulation fraction as a function of  $\delta$ assuming v$_{earth}$=227$\pm$14.4 km/s. The above plots assumed galactic escape velocity v$_{escape}$=500 km/s.}}
  \label{idm_dru} 
\end{center}
\end{figure} 
\begin{figure*}[!htb]
\begin{center}
$\begin{array}{c@{\hspace{.01in}}c@{\hspace{.01in}}c}
\includegraphics[width=3in]{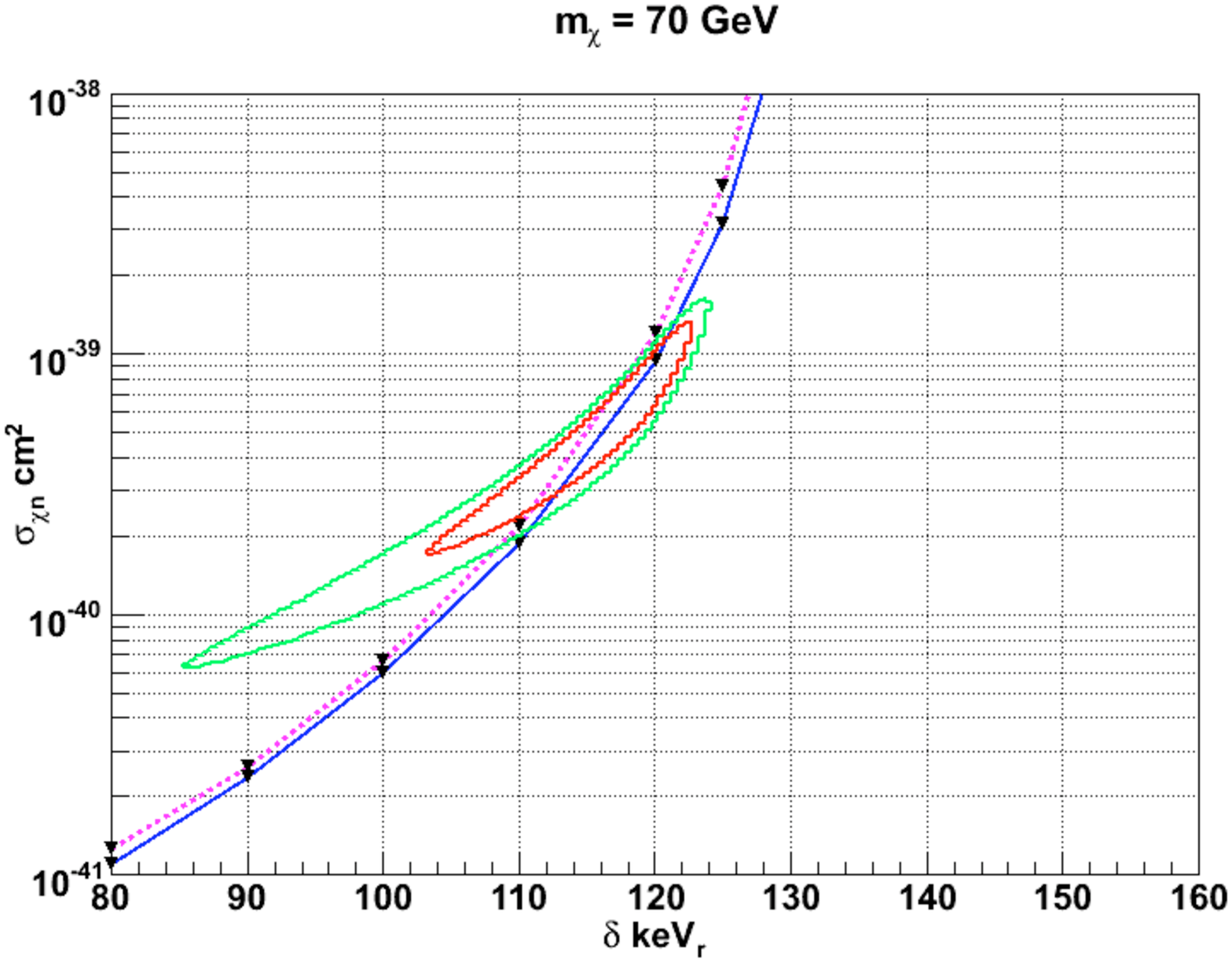}  &  
\includegraphics[width=3in]{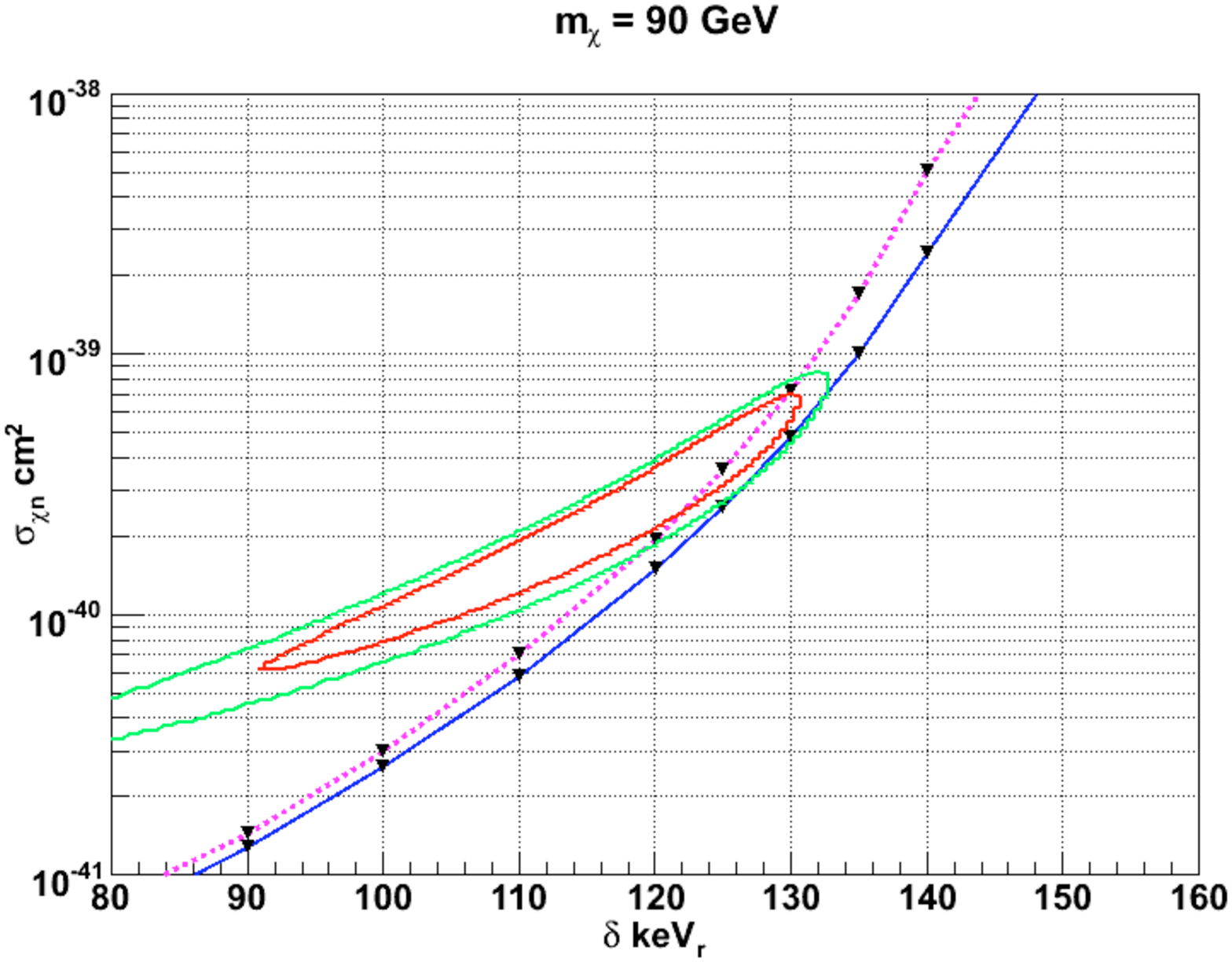} \\
\includegraphics[width=3in]{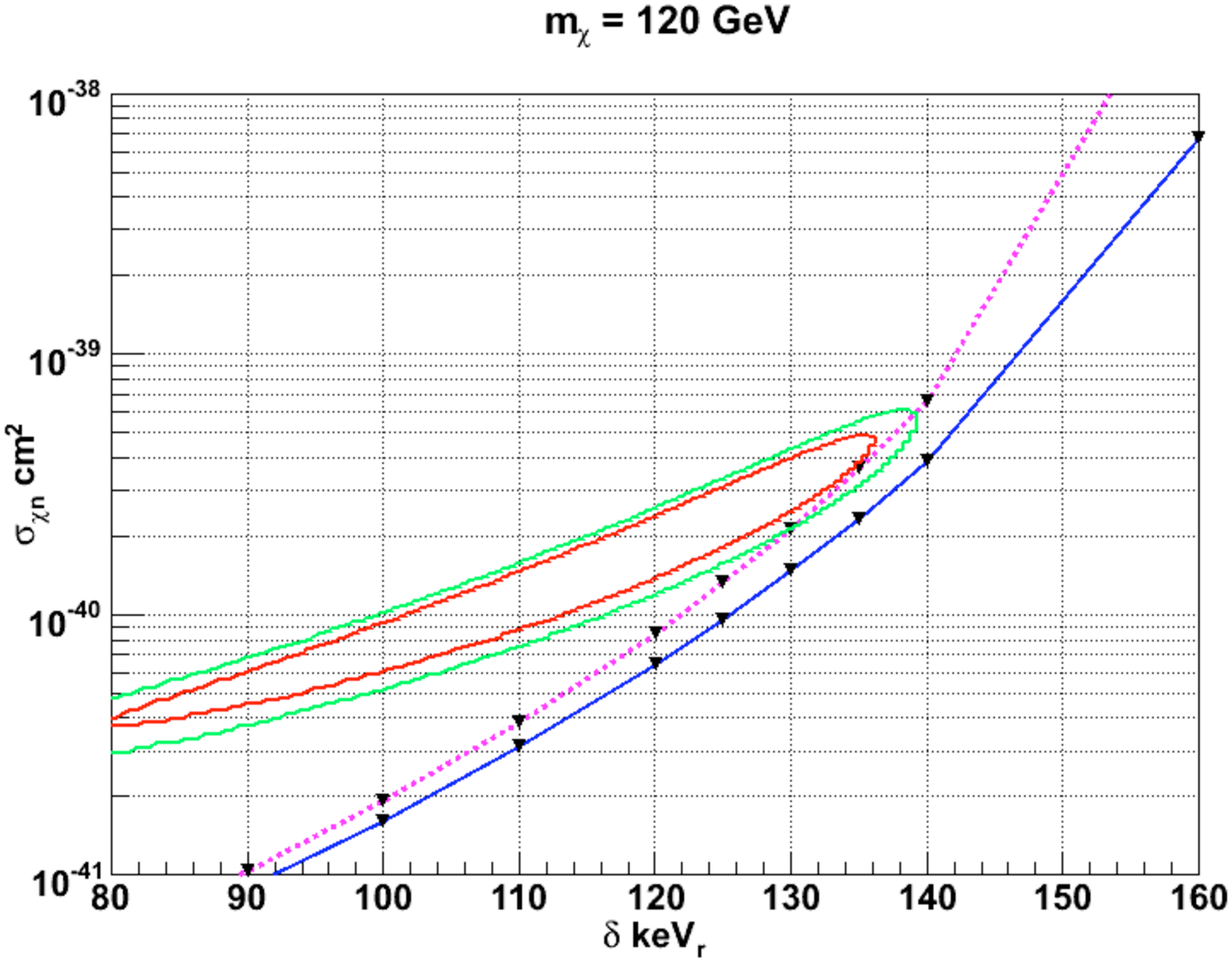} &
\includegraphics[width=3in]{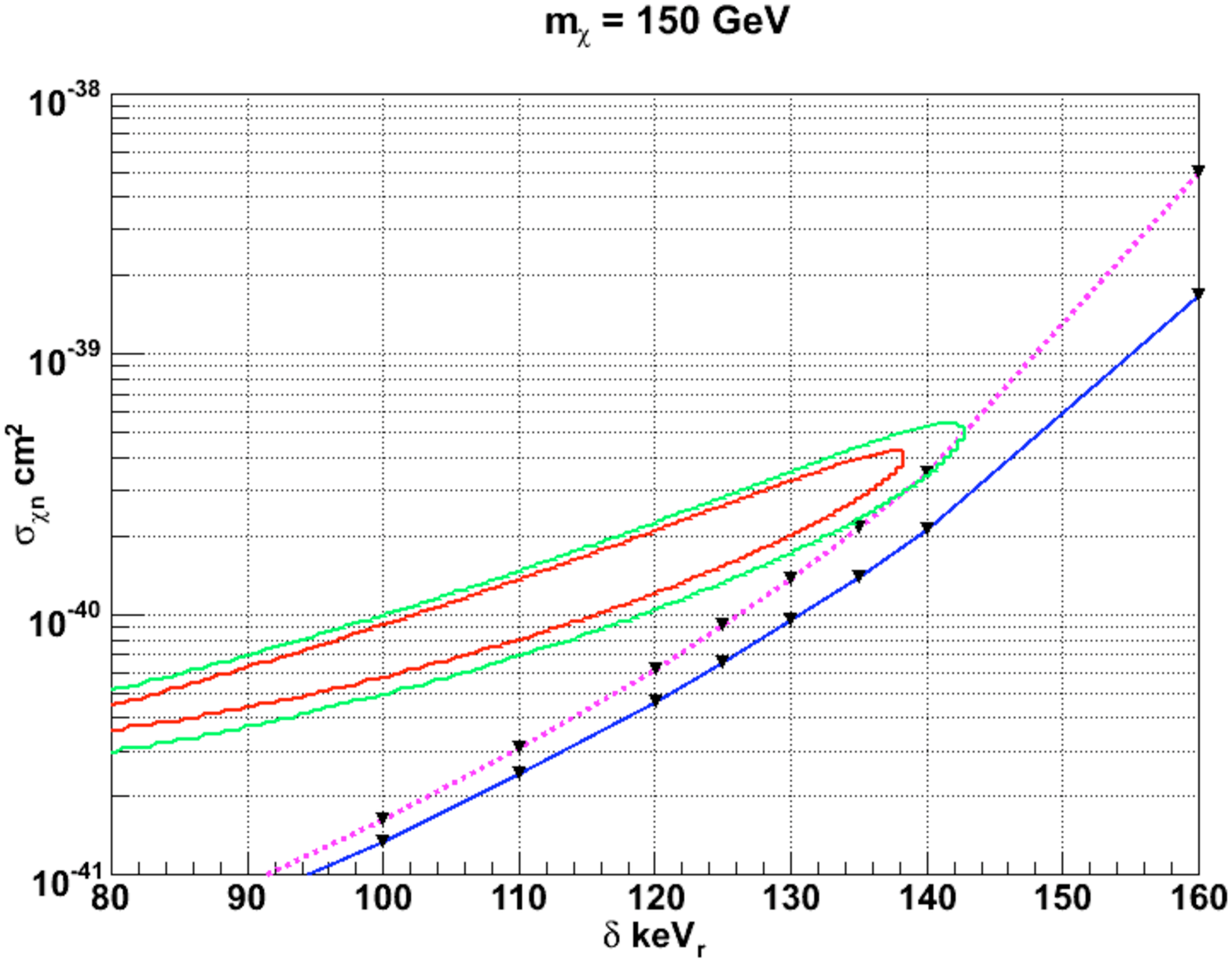} \\
\includegraphics[width=3in]{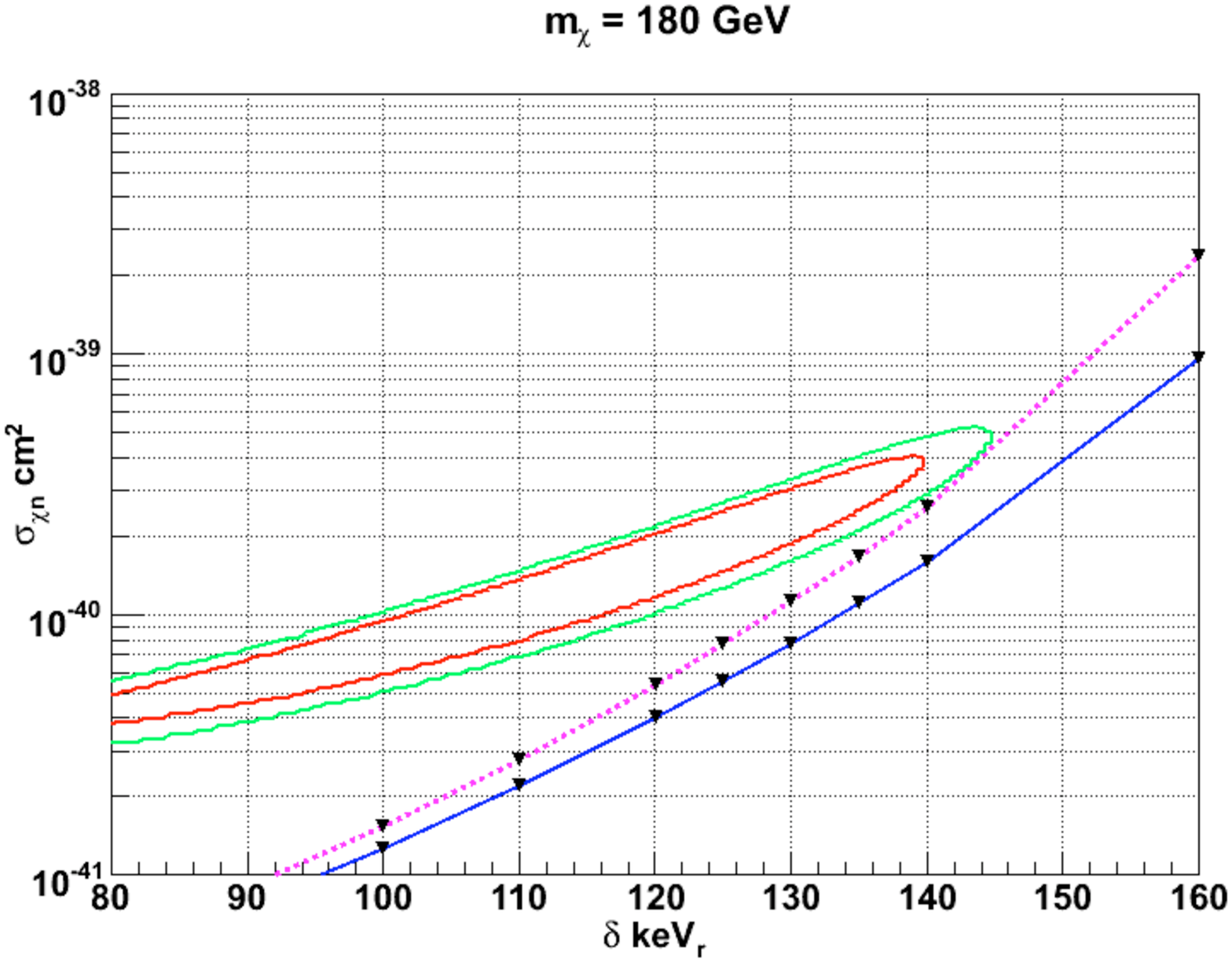} &
\includegraphics[width=3in]{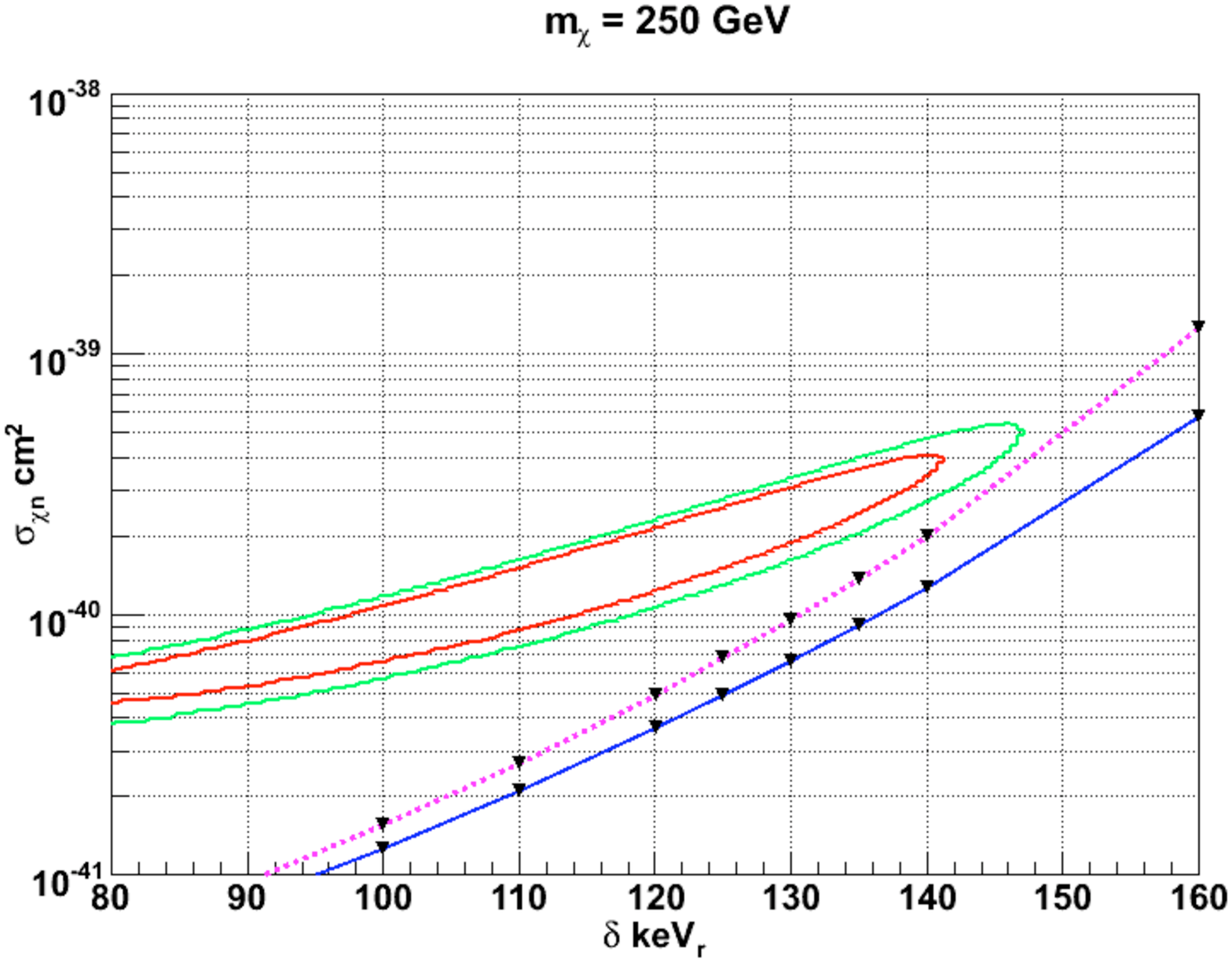} \\
\end{array}$
\caption{\footnotesize{Preferred regions for DAMA/LIBRA results and ZEPLIN-II exclusion limits. The blue solid lines are ZEPLIN-II exclusion limits at 90\% C.L., assuming xenon zero-field quenching factor qXe=0.19. The pink dotted lines are the exclusion limits assuming the most recent qXe reported in Ref.~\cite{sorensen}. The DAMA/LIBRA contours are at 90\%(red) and 99\% (green) C.L. with standard qI=0.085, following the analysis done in Ref.~\cite{weiner09}. These plots assumed $v_{escape}$= 500 km/s.}}
 \label{zepii_1_dama} 
\end{center}
\end{figure*}
The inelastic dark matter (iDM)~\cite{weiner09}, ~\cite{weiner05},~\cite{weiner01} model explored the possibility of an altered kinematic of the WIMP-nucleus interaction in an attempt to reconcile the detection of  annual modulation observation~\cite{dama08} and null results from all other experiments such as CDMS~\cite{cdms2009}, XENON10~\cite{xenon10} and ZEPLIN-II~\cite{zepsi}.  iDM assumes two basic properties: a) WIMP $\chi$ has an excited state $\chi$*, with a mass m$_{\chi^*}$$-$m$_{\chi}$ = $\delta \sim$ 100 keV recoil-energy (keV$_r$). b) Elastic scatterings of $\chi$ off target nucleus N are suppressed compared with the inelastic scattering $\chi$N$\rightarrow$$\chi^*$N. Refer to Ref.~\cite{weiner01}, ~\cite{russell} and ~\cite{randall} for discussions on particle models that may give rise to this scenario. \\
Compared to the elastic case, the introduction of splitting $\delta$ in the scattering kinematics increases the minimum velocity $\beta_{min}$ of $\chi$ to scatter with a deposited recoil energy  E$_R$. The new $\beta_{min}$ can be shown to be\\
\begin{equation}
\beta_{min}=\sqrt{\frac{1}{2m_NE_R}}\Biggl(\frac{m_NE_R}{\mu}+\delta\Biggr)
\end{equation}
\\
where m$_N$ is the mass of the target nucleus and $\mu$ is the reduced mass of the $\chi$/target nucleus system.
For direct detection experiments, the key consequences of iDM are:
\begin{enumerate}
\item	Differential event rate is lowered since minimum velocity $\beta_{min}$ is increased for a given recoil energy.  
\item	Low-energy events are suppressed in the spectrum of events. In the elastic case, event rate is the highest in the low recoil energy, while the splitting $\delta \sim$100 keV$_r$ in iDM scenario can suppress or even eliminate low energy events. 
\item	Annual modulation of signal is enhanced. In the extreme case, where the $\chi$ particles only have enough minimum velocity in the summer but not in the winter, the modulation can be 100\%, albeit at a significant lower overall event rate.
\item	Heavier target nuclei are even more favored. While it is true that in the elastic case heavier target nuclei gives rises to a higher differential event rate, inelastic scattering favors heavier target more since $\beta_{min}$ is lowered significantly for heavier elements.  
\end{enumerate} 
 These key characteristics are quantified in Fig~\ref{idm_dru}. An effective test to the WIMPs discovery signals reported in Ref.~\cite{dama08} with Iodine (I) target (A=127) is to use a different target with comparable atomic mass. ZEPLIN-II with Xe (A=132) target is a good experiment for this test. 

\section{ZEPLIN-II  Inelastic Dark Matter Limits}
The first ZEPLIN-II  dark matter run recorded 29 events in the acceptance box with an expected background of 28.6 $\pm$ 4.3. The result is consistent with a null experiment with an upper limit of 10.4 using Feldman Cousins method~\cite{cousins}. We maintained that the number of observed events in the acceptance box is consistent with the tail distributions from the two background populations. On the other hand, a confidence interval derived by taking all 29 events as iDM signals without background subtractions, such as the p$_{Max}$ method detailed in Ref.~\cite{yellin}, may not be the best approach to interpret ZEPLIN-II results since p$_{Max}$ is in general more suitable for experiments with small number of events where a statistical characterization of background is more challenging. We maintain that we understood our background accurately enough to use Feldman Cousins' method, which has an additional advantage of guaranteed correct coverage.\\
Fig.~\ref{zepii_1_dama} shows iDM ZEPLIN-II exclusion limits and DAMA signal contours for different WIMP masses. DAMA contours of 90\% (red, $\bigtriangleup \chi^2$=6.25) and 99\% (green, $\bigtriangleup \chi^2$=11.34) C.L.  were constructed following the treatments by Ref.~\cite{weiner09}  using iodine quenching factor (qI) of 0.085, where E$_{R}$ = E$_{ee}$/qI. For ZEPLIN-II, the zero field quenching factor qXe is modified by the presence of the electric field, and the conversion equation becomes E$_{R}$ = E$_{ee}$/qXe$\cdot$(Se/Sn), where the scintillation quenching of electron and nuclear recoils due to the electric field are S$_e$=0.54$\pm$0.02~\cite{se} and S$_n$=0.95$\pm$0.05~\cite{sn}. Assuming Standard Halo Model (SHM) parameters outlined in Ref.~\cite{peter_limit} and ~\cite{savage}, and a constant qXe of 0.19 (giving E$_{R}$ = E$_{ee}$/0.36), the resultant ZEPLIN-II inelastic dark matter limits are shown in blue solid lines, for the WIMP masses between 70 GeV $\leq$ m$_{\chi}$ $\leq$ 250 GeV. Using these parameters, it can be seen that these limits constrained a larger iDM parameter space than those previously reported in Ref.~\cite{weiner09}, suggesting the exclusion of published claims for iDM signals at $>$99\% C.L. for WIMP masses $>$100 GeV. Also shown in pink dotted lines are ZEPLIN-II limits assuming the the most recent energy dependent qXe values reported in Ref.~\cite{sorensen}, which estimated an increase in these values at high recoil energies (qXe $\sim$0.24 at 50 keV$_r$). The new  qXe measurements lowered the exclusion power of ZEPLIN-II results, but at WIMPs masses $>$180 GeV there are still strong indications of exclusion of discovery claims at $>$99\% C.L. Incidentally, uncertainty in qI, which is reported as qI=0.09$\pm$0.01~\cite{que_iodine}, has a significant effect on DAMA/LIBRA preferred regions in iDM case. In general, a higher qI would increase the discrepancy between DAMA/LIBRA and the other experiments, while a lower qI would lead to a better agreement. Refer to Ref.~\cite{russell} for a more detailed discussion of the effects of qI, as well as other cosmological parameters, on the constraints of iDM.

%
\section{Conclusion}
In conclusion, the published ZEPLIN-II background analysis reported in Ref.~\cite{zepsi} showed that the events observed in the acceptance box were statistically consistent with the tail distributions from two background populations; namely the $\gamma$ and radon-induced events. As such it is natural to take the background subtraction approach to analyze the events in the acceptance region, from which we declared ZEPLIN-II a null experiment with a 90\% C.L. upper limit on WIMP signal number at 10.4. This translated to iDM exclusion limits that are significantly more stringent than the previously published values in Ref.~\cite{weiner09}, which adopted p$_{Max}$ method to derive an upper limit. Using parameters from Standard Halo Model,  ZEPLIN-II limits suggest the exclusion of published claims for iDM signals at $>$99\% C.L., for WIMP masses $>$100 GeV.\\\\
This work has been funded by the US Department of Energy (grant numbers
DE-FG03-91ER40662 and DE-FG03-95ER40917) and the 
US National Science Foundation (grant number 
PHY-0139065 and PHY-06-53459). We would like to thank members of ZEPLIN-II Collaboration for the help and support. We are also indebted to N. Weiner for valuable advice on the details of iDM model and we acknowledge helpful discussions with P.F. Smith. D.B. Cline would like to thank the Aspen Institute for Physics where part of this paper was completed.   

%

\end{document}